\documentclass[runningheads]{llncs}
\usepackage[utf8]{inputenc}
\usepackage{textcomp}
\usepackage{graphicx,verbatim}
\usepackage{hyperref}
\usepackage{cite}
\usepackage{amsmath}
\usepackage{amssymb}
\usepackage{authblk}
\usepackage{bbding}
\usepackage{graphicx}
\usepackage{booktabs}
\usepackage{graphicx}
\usepackage[normalem]{ulem}
\usepackage{multirow}
\usepackage{amsmath} 
\usepackage{bm}      
\usepackage{multirow}
\usepackage[table,xcdraw]{xcolor}
\usepackage{colortbl}
\usepackage[normalem]{ulem}
\useunder{\uline}{\ul}{}
\usepackage[ruled,vlined]{algorithm2e}
\usepackage{algpseudocode} 

\begin{document}
\title{WeGA: Weakly-Supervised Global-Local Affinity Learning Framework for Lymph Node Metastasis Prediction in Rectal Cancer}
\titlerunning{Weakly-Supervised Global-Local Affinity Learning}

\author{Yifan Gao\inst{1,2,3}\textsuperscript{*}, Yaoxian Dong\inst{1,2}\textsuperscript{*} , Wenbin Wu\inst{1,2}, Chaoyang Ge\inst{1,2}, Feng Yuan\inst{1,2}, Jiaxi Sheng\inst{1,2}, Haoyue Li\inst{1,4}, Xin Gao\inst{2}\textsuperscript{\Envelope} }  
\authorrunning{Y. Gao et al.}
\institute{
    School of Biomedical Engineering (Suzhou), Division of Life Science and Medicine, University of Science and Technology of China, Hefei, China \and
    Suzhou Institute of Biomedical Engineering and Technology, Chinese Academy of Sciences, Suzhou, China \and
    Shanghai Innovation Institute, Shanghai, China \and
    College of Medicine and Biological Information Engineering, Northeastern University, Shenyang, China \\
}

\maketitle              

\begingroup
\renewcommand\thefootnote{\textit{*}}
\footnotetext{These authors contributed equally to this work.}
\endgroup

\begin{abstract}
Accurate lymph node metastasis (LNM) assessment in rectal cancer is essential for treatment planning, yet current MRI-based evaluation shows unsatisfactory accuracy, leading to suboptimal clinical decisions. Developing automated systems also faces significant obstacles, primarily the lack of node-level annotations. Previous methods treat lymph nodes as isolated entities rather than as an interconnected system, overlooking valuable spatial and contextual information. To solve this problem, we present WeGA, a novel weakly-supervised global-local affinity learning framework that addresses these challenges through three key innovations: 1) a dual-branch architecture with DINOv2 backbone for global context and residual encoder for local node details; 2) a global-local affinity extractor that aligns features across scales through cross-attention fusion; and 3) a regional affinity loss that enforces structural coherence between classification maps and anatomical regions. Experiments across one internal and two external test centers demonstrate that WeGA outperforms existing methods, achieving AUCs of 0.750, 0.822, and 0.802 respectively. By effectively modeling the relationships between individual lymph nodes and their collective context, WeGA provides a more accurate and generalizable approach for lymph node metastasis prediction, potentially enhancing diagnostic precision and treatment selection for rectal cancer patients.

\end{abstract}

\section{Introduction}
Rectal cancer treatment planning hinges on accurate lymph node metastasis (LNM) assessment, with significant implications for patient outcomes \cite{glynne2017rectal,cohen2021combining,zwager2022deep}. Patients without LNM may benefit from less invasive local excision, while those with nodal involvement require more aggressive approaches including neoadjuvant chemoradiotherapy followed by radical surgery \cite{liu2021total}. Although it plays a crucial role in determining treatment pathways, the accuracy of lymph node assessment through Magnetic Resonance Imaging (MRI) remains unsatisfactory \cite{blazic2016mri,horvat2019mri}. This suboptimal performance stems from the similar appearance of benign and metastatic lymph nodes in conventional imaging \cite{brouwer2018clinical,grone2018accuracy}.

Driven by the rapid development of artificial intelligence \cite{dai2021transmed,gao2023anatomy,gao2024desam,gao2024mba}, automated lymph node evaluation methods show significant promise \cite{yang2024sagl,yang2021high,kudo2021artificial,abbaspour2024application}, but simultaneously faces considerable challenges, primarily due to the lack of annotations at the lymph node level. Matching individual lymph nodes on preoperative MRI to postoperative pathology specimens proves exceptionally difficult, especially for small nodes.

While weakly-supervised learning offers a promising direction by utilizing patient-level labels to guide node-level predictions, current approaches have yet to effectively model the intricate relationships that exist within the lymphatic system \cite{xia2024multicenter}. Specifically, they treat lymph nodes as isolated entities rather than as interconnected components within a larger biological context. These approaches fail to capture the spatial relationships between nodes, which contain valuable diagnostic information that experienced radiologists use in their assessments. This raises an important question: "\textbf{Can we design a learning paradigm that effectively bridges the gap between patient-level supervision and node-level predictions?}"

In order to solve this problem, we present WeGA, a novel weakly-supervised global-local affinity learning framework for lymph node metastasis prediction. WeGA employs a dual-branch architecture with DINOv2 \cite{oquab2024dinov} backbone for processing global lymphatic context and a ResNet encoder for capturing local node details. We introduce a global-local affinity learning extractor that aligns information across scales through cross-attention fusion, connecting individual node characteristics with their anatomical context. Furthermore, we develop a regional affinity loss that enhances weakly-supervised learning by enforcing structural coherence between classification maps and anatomical regions, working alongside multi-Instance learning and label proportion learning objectives.

\begin{algorithm}[t!]
\caption{Overview of our proposed WeGA framework}
\label{alg:wega}
\SetKwInOut{Input}{Input}
\SetKwInOut{Output}{Output}

\Input{Local\_patch: 32×32 LN sub-image \\
Global\_patchwork: 128×128 stitched LN image \\
LN features: $\mathbf{f}_{\text{size}}^j$, $\mathbf{f}_{\text{SD}}^j$, $\mathbf{f}_{\text{RD}}^j$, $\mathbf{f}_{\text{ADC}}^j$ \\
}
\Output{$\hat{y}_p$: Patient-level prediction, $\{p^j\}$: LN probabilities}

\textbf{Global-Local Affinity Learning:}
\For{each patient $i$}{
\textbf{Global Branch:} \\
1. Process composite image $\mathcal{G}_i$ through pre-trained DINOv2 backbone: \\
\quad $\mathbf{H}_g = \text{DINOv2}(\mathcal{G}_i)$ \\
2. Retain multi-scale features: \\
\quad $\{\mathbf{h}_g^{(1)}, \mathbf{h}_g^{(5)}, \mathbf{h}_g^{(9)}\} \leftarrow \text{TransformerBlock}_{[1,5,9]}(\mathbf{H}_g)$ \\
  
\textbf{Local Branch:} \\
\For{each LN patch $\mathcal{P}_i^j$}{
1. Process through ResNet-like encoder: \\
\quad $\mathbf{F}_l^j = \text{ResNetLayers}(\mathcal{P}_i^j)$ \\
2. \textbf{Global-local Affinity Extractor:} \\
\quad a. Embed features into tokens: $\mathbf{E}_l^j = \text{Embed}(\mathbf{F}_l^j)$ \\
\For{$k \in \{1,5,9\}$}{
\quad b. Apply cross-attention: $\mathbf{C}_k^j = \text{CrossAttn}_k(\mathbf{E}_l^j, \mathbf{h}_g^{(k)})$ \\
}
\quad c. Convert tokens back to spatial features: $\mathbf{F}_c^j = \text{Unembed}(\mathbf{C}_k^j)$ \\
3. \textbf{Metastasis Prediction:} \\
\quad $p^j = \sigma(\text{MLP}_c(\mathbf{F}_c^j \oplus \mathbf{f}_{\text{size}}^j \oplus \mathbf{f}_{\text{SD}}^j \oplus \mathbf{f}_{\text{RD}}^j \oplus \mathbf{f}_{\text{ADC}}^j))$
}
}

\textbf{Weakly Supervised Optimization:} \\
1. Compute $\mathcal{L}_{\text{MIL}}$, $\mathcal{L}_{\text{LLP}}$, and $\mathcal{L}_{\text{RA}}$ as defined in Equations (5), (6), and (10) \\
2. Update model parameters using $\mathcal{L}_{\text{total}} = \alpha\mathcal{L}_{\text{MIL}} + \beta\mathcal{L}_{\text{LLP}} + \gamma\mathcal{L}_{\text{RA}}$ \\
\end{algorithm}

Our experimental results demonstrate that WeGA achieves superior performance in identifying lymph node metastasis across diverse clinical settings. Unlike previous approaches that focus solely on isolated node features, our method effectively models the interrelationships between individual lymph nodes and their collective context, providing a more comprehensive assessment of metastatic status and potentially improving diagnostic accuracy and treatment planning for rectal cancer patients.

\begin{figure}
	\includegraphics[width=\textwidth]{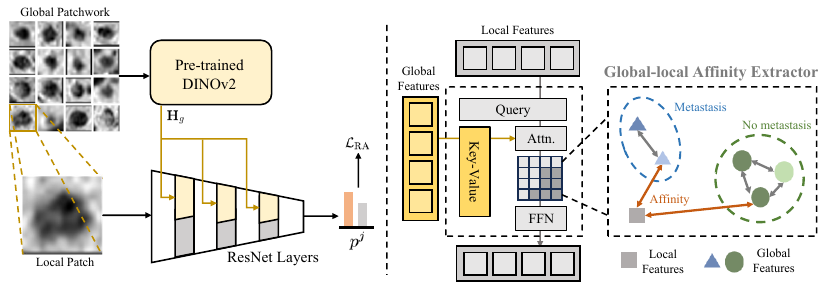}
	\centering
	\caption{Overview of the WeGA framework. The dual-branch design processes both global context and local node details, followed by global-local affinity extractor to align multi-scale features for comprehensive representation learning.} 
	\label{fig1}
\end{figure}

\section{Methodology}
The complete procedure for our WeGA framework is outlined in Algorithm~\ref{alg:wega}. Our combined network architecture integrates a main network for processing individual lymph node patches and an auxiliary network for analyzing the composite lymph node image. The auxiliary network extracts multi-scale features from the DINOv2 backbone, which are then used by the main network to establish global-local affinities through cross-attention mechanisms. Moreover, we proposed regional affinity loss to effectively addresses the constraints of weakly-supervised learning by combining complementary loss functions that operate at different levels of granularity.

Our proposed WeGA framework builds upon the \cite{xia2024multicenter} method for lymph node metastasis prediction in rectal cancer. While previous method provides a foundation with radiomics feature extraction and initial weakly-supervised learning approaches, WeGA introduces several key innovations: a dual-branch architecture capturing both global and local contexts, a global-local affinity extractor based on cross-attention, and a regional affinity loss to enhance localization consistency. These innovations collectively improve lymph node metastasis prediction with only patient-level annotations.

\textbf{Network Architecture: }Accurate lymph node metastasis prediction requires analyzing both node-specific details and their contextual relationships. Our WeGA framework introduces a dual-branch architecture that processes both scales simultaneously, as illustrated in Fig. \ref{fig1}.

The architecture consists of a global branch and a local branch. For each patient, we extract individual lymph node patches $\{\mathcal{P}_i^j\}$ and create a composite image $\mathcal{G}_i$ by stitching all lymph node regions together.

The global branch employs DINOv2 to process the composite image: $\mathbf{H}_g = \text{DINOv2}(\mathcal{G}_i)$, where $\mathbf{H}_g$ represents the learned embeddings. We extract multi-scale features from different transformer blocks:

\begin{equation}
\{\mathbf{h}_g^{(1)}, \mathbf{h}_g^{(5)}, \mathbf{h}_g^{(9)}\} \leftarrow \text{TransformerBlock}_{[1,5,9]}(\mathbf{H}_g)
\end{equation}

The local branch processes each lymph node patch through a ResNet-inspired network consisting of specialized BasicBlocks, $\mathbf{F}_l^j = \text{ResNetLayers}(\mathcal{P}_i^j)
$, where $\mathbf{F}_l^j$ represents spatial features extracted from the ResNet-like network. This branch includes several convolutional layers followed by BasicBlocks that extract increasingly abstract representations.

This dual-branch design provides a foundation for connecting patient-level supervision with node-level predictions through the global-local affinity extractor described in the next section.

\textbf{Global-Local Affinity Extractor: }The effective transfer of knowledge from patient-level annotations to node-level predictions requires a mechanism to establish meaningful connections between global contextual information and local node-specific features. Our affinity extractor addresses this necessity by integrating complementary feature representations through cross-attention.

The local branch features from the ResNet-like architecture are first transformed into token embeddings through a patch-based embedding process:

\begin{equation}
\mathbf{E}_l^j = \text{Embed}(\mathbf{F}_l^j, \text{patch\_size})
\end{equation}

\noindent where $\text{Embed}$ transforms spatial features into a sequence of patch tokens with an additional classification token.

The core of our proposed module is a cross-attention mechanism that aligns and integrates local and global representations. For each scale $k \in \{1,5,9\}$, we compute: $\mathbf{C}_k^j = \text{CrossAttn}_k(\mathbf{E}_l^j, \mathbf{h}_g^{(k)})$, where $\text{CrossAttn}_k$ implements transformer blocks with cross-attention. This operation allows local lymph node features to attend to relevant global contextual cues. The cross-attention computes compatibility scores between local queries and global keys:

\begin{equation}
\text{Attention}(\mathbf{Q}, \mathbf{K}, \mathbf{V}) = \text{softmax}\left(\frac{\mathbf{Q}\mathbf{K}^T}{\sqrt{d}}\right)\mathbf{V}
\end{equation}

\noindent where $\mathbf{Q}$ is derived from local features, while $\mathbf{K}$ and $\mathbf{V}$ are derived from global features at scale $k$.

After cross-attention, the token embeddings are converted back to spatial representations through an unembedding process: $\mathbf{F}_c^j = \text{Unembed}(\mathbf{C}_k^j)$, where $\text{Unembed}$ transforms token embeddings back to spatial feature maps through reshaping and transposed convolution operations.

This cross-attention block establishes relationships between individual lymph nodes and the patient's overall lymph node distribution, facilitating knowledge transfer from patient-level supervision to node-level predictions. The final lymph node representation integrates these cross-attention features with radiomics features to produce a comprehensive feature set for metastasis prediction:

\begin{equation}
p^j = \sigma(\text{MLP}_c(\mathbf{F}_c^j \oplus \mathbf{f}_{\text{size}}^j \oplus \mathbf{f}_{\text{SD}}^j \oplus \mathbf{f}_{\text{RD}}^j \oplus \mathbf{f}_{\text{ADC}}^j))
\end{equation}

\noindent where $\oplus$ denotes feature concatenation, and $\text{MLP}_c$ represents a multi-layer perceptron for classification.

Through this affinity extractor, our WeGA framework effectively distributes supervisory signals from patient-level labels to node-level predictions, enabling accurate lymph node metastasis detection despite the constraints of weakly-supervised learning.

\textbf{Loss Functions: } Our WeGA framework employs multiple loss functions to optimize the model under the weakly-supervised setting. We retain two loss functions from \cite{xia2024multicenter}, and introduce a new regional affinity loss to enhance localization consistency.

The Multi-Instance Loss ($\mathcal{L}_{\text{MIL}}$) leverages the assumption that if a patient has metastasis, at least one lymph node must be metastatic. Conversely, if a patient is metastasis-free, all lymph nodes must be non-metastatic:

\begin{equation}
\mathcal{L}_{\text{MIL}} = -\sum_i y_i\log(\max_j p_i^j) + (1-y_i)\log(1-\max_j p_i^j)
\end{equation}

The Label Proportion Loss ($\mathcal{L}_{\text{LLP}}$) incorporates prior knowledge about the proportion of metastatic lymph nodes in each patient:

\begin{equation}
\mathcal{L}_{\text{LLP}} = \frac{1}{N}\sum_i \|\frac{1}{|\mathcal{P}_i|}\sum_j p_i^j - \frac{m_i}{M_i}\|_2^2
\end{equation}

\noindent where $\frac{m_i}{M_i}$ represents the ratio of the estimated number of metastatic nodes ($m_i$) to the total number of nodes ($M_i$) for patient $i$.

To enhance localization consistency under weak supervision, we propose a variance-guided regional affinity loss (RAL). This regularization term leverages the relationship between prediction confidence and feature stability. 

Let $\mathbf{z} \in \mathbb{R}^{N\times C\times H\times W}$ denote the logits from the final convolutional layer. We first compute class probabilities through softmax activation:

\begin{equation}
p_{n,c,h,w} = \frac{e^{z_{n,c,h,w}}}{\sum_{c'=1}^C e^{z_{n,c',h,w}}}
\end{equation}

Spatial variance maps are generated by measuring prediction uncertainty across classes:

\begin{figure}[t!]
	\includegraphics[width=1.0\textwidth]{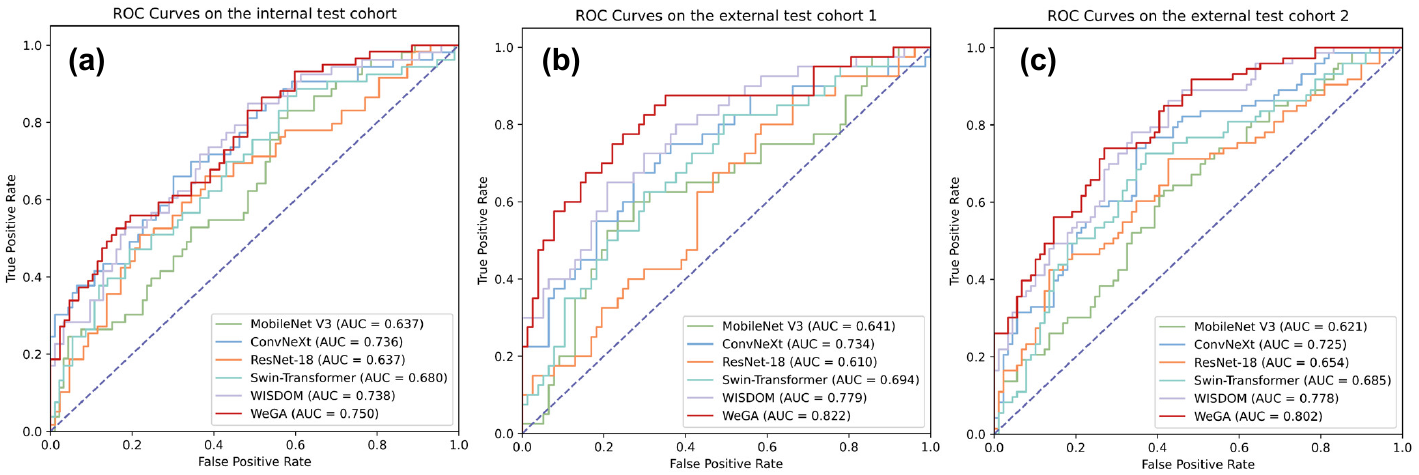}
	\centering
	\caption{ROC curves comparing the proposed WeGA framework with baseline methods across three independent test centers.} 
	\label{fig2}
\end{figure}

\begin{equation}
\sigma^2_{n,h,w} = \frac{1}{C}\sum_{c=1}^C \left(p_{n,c,h,w} - \frac{1}{C}\sum_{c'=1}^C p_{n,c',h,w}\right)^2
\end{equation}

Normalized variance $\tilde{\sigma}^2_{n,h,w}$ is scaled to [0,1] through min-max normalization over all spatial positions. Dynamic masks are then defined as:

\begin{equation}
\mathbf{M}_{\text{bg}} = \mathbb{I}(\tilde{\sigma}^2 < \theta_{\text{bg}}), \quad \mathbf{M}_{\text{fg}} = \mathbb{I}(\tilde{\sigma}^2 > \theta_{\text{bg}} + \Delta_{\text{bg-fg}})
\end{equation}

\noindent where $\mathbb{I}(\cdot)$ is the indicator function, $\theta_{\text{bg}}$ thresholds background regions, and $\Delta_{\text{bg-fg}}$ ensures separation between foreground/background. The regularization term penalizes conflicting predictions:

\begin{equation}
\mathcal{L}_{\text{RA}} = \frac{1}{|\Omega|}\sum_{(n,h,w)\in\Omega} \left[\sigma(z_{n,y_n,h,w})\cdot\mathbf{M}_{\text{bg}} + (1-\sigma(z_{n,y_n,h,w}))\cdot\mathbf{M}_{\text{fg}}\right]
\end{equation}

This complements existing MIL and LLP losses by enforcing spatial coherence: background regions with stable features suppress false activations, while high-variance areas amplify metastasis-related responses. The thresholds $\theta_{\text{bg}}$ and $\Delta_{\text{bg-fg}}$ are empirically set to 0.3 and 0.1 respectively, balancing sensitivity and specificity across multi-center data.

The total loss function is a weighted combination of these three components:

\begin{equation}
\mathcal{L}_{\text{total}} = \alpha\mathcal{L}_{\text{MIL}} + \beta\mathcal{L}_{\text{LLP}} + \gamma\mathcal{L}_{\text{RA}}
\end{equation}

\noindent where $\alpha=1$, $\beta=0.5$, and $\gamma=0.5$ are hyperparameters that control the contribution of each loss component.

\begin{figure}
	\includegraphics[width=1.0\textwidth]{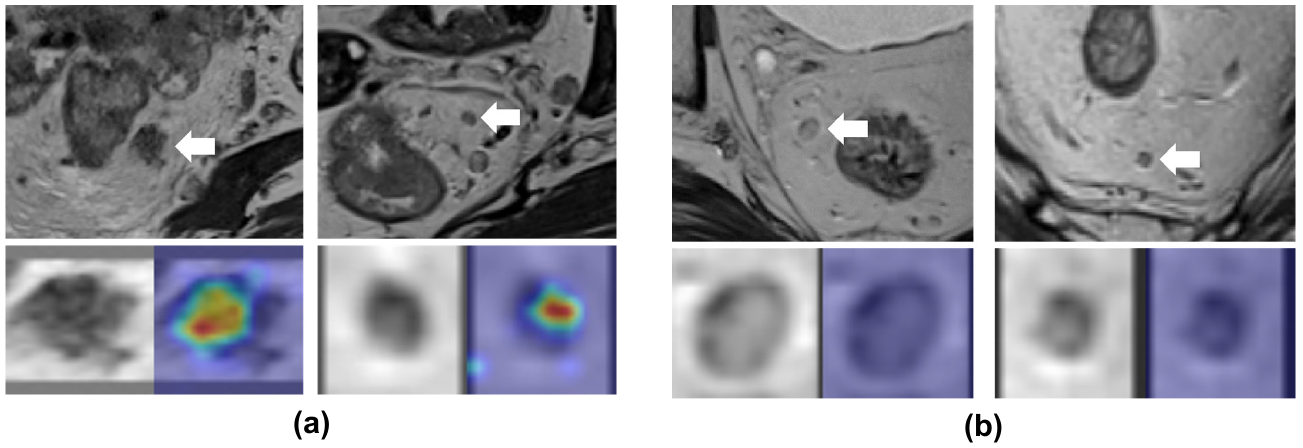}
	\centering
	\caption{Heatmap visualization of our proposed framework. (a) Lymph node imaging of patients with metastasis. (b) Lymph node imaging of patients without metastasis.} 
	\label{fig3}
\end{figure}

\section{Experiments and Results}

\textbf{Implementation and Experiment Setting: } We implemented the WeGA framework using PyTorch and conducted evaluations across multiple medical centers. The experiments utilized T2-weighted MRI scans from rectal cancer patients collected from three centers with institutional review board approval. We divided data from one center in a 7:1:2 ratio, resulting in an internal test center (ITC) dataset with 147 patients. The remaining two centers served as external validation cohorts (ETC1 and ETC2) with 117 and 162 patients, respectively. 

\begin{table}[]
\centering
\caption{Comparison of the proposed WeGA framework with state-of-the-art methods for lymph node metastasis prediction in rectal cancer. Performance metrics include Area Under the Curve (AUC), Accuracy (ACC), and F1 score evaluated on one internal test center (ITC) and two external test centers (ETC1, ETC2). Bold values indicate the best performance for each metric and test center.}
\label{tab:0}
\resizebox{\columnwidth}{!}{%
\begin{tabular}{c|ccc|ccc|ccc}
\toprule
\multirow{2}{*}{Model}            & \multicolumn{3}{c|}{AUC}                                                                                                                                                                                                                             & \multicolumn{3}{c|}{ACC}                                                                                                                                                                                                                             & \multicolumn{3}{c}{F1}                                                                                                                                                                                                                              \\ 
                                  & ITC                                                                             & ETC1                                                                            & ETC2                                                                            & ITC                                                                             & ETC1                                                                            & ETC2                                                                            & ITC                                                                             & ETC1                                                                            & ETC2                                                                            \\ \midrule
\multirow{2}{*}{Mobilenet V3 \cite{howard2019searching}}     & \multirow{2}{*}{\begin{tabular}[c]{@{}c@{}}0.637\\ (0.544, 0.727)\end{tabular}} & \multirow{2}{*}{\begin{tabular}[c]{@{}c@{}}0.641\\ (0.527, 0.752)\end{tabular}} & \multirow{2}{*}{\begin{tabular}[c]{@{}c@{}}0.621\\ (0.53, 0.708)\end{tabular}}  & \multirow{2}{*}{\begin{tabular}[c]{@{}c@{}}0.568\\ (0.486, 0.644)\end{tabular}} & \multirow{2}{*}{\begin{tabular}[c]{@{}c@{}}0.676\\ (0.59, 0.761)\end{tabular}}  & \multirow{2}{*}{\begin{tabular}[c]{@{}c@{}}0.598\\ (0.519, 0.673)\end{tabular}} & \multirow{2}{*}{\begin{tabular}[c]{@{}c@{}}0.568\\ (0468, 0.662)\end{tabular}}           & \multirow{2}{*}{\begin{tabular}[c]{@{}c@{}}0.546\\ (0.410, 0.667)\end{tabular}} & \multirow{2}{*}{\begin{tabular}[c]{@{}c@{}}0.578\\ (0.478, 0.667)\end{tabular}} \\
                                  &                                                                                 &                                                                                 &                                                                                 &                                                                                 &                                                                                 &                                                                                 &                                                                                          &                                                                                 &                                                                                 \\
\multirow{2}{*}{ConvNeXt \cite{woo2023convnext}}         & \multirow{2}{*}{\begin{tabular}[c]{@{}c@{}}0.736\\ (0.644, 0.817)\end{tabular}} & \multirow{2}{*}{\begin{tabular}[c]{@{}c@{}}0.734\\ (0.627, 0.829)\end{tabular}} & \multirow{2}{*}{\begin{tabular}[c]{@{}c@{}}0.725\\ (0.643, 0.800)\end{tabular}} & \multirow{2}{*}{\begin{tabular}[c]{@{}c@{}}0.677\\ (0.596, 0.753)\end{tabular}} & \multirow{2}{*}{\begin{tabular}[c]{@{}c@{}}0.702\\ (0.615, 0.778)\end{tabular}} & \multirow{2}{*}{\begin{tabular}[c]{@{}c@{}}0.686\\ (0.611, 0.759)\end{tabular}} & \multirow{2}{*}{\begin{tabular}[c]{@{}c@{}}0.587\\ (0.472, 0.688)\end{tabular}}          & \multirow{2}{*}{\begin{tabular}[c]{@{}c@{}}0.597\\ (0.463, 0.709)\end{tabular}} & \multirow{2}{*}{\begin{tabular}[c]{@{}c@{}}0.675\\ (0.584, 0.756)\end{tabular}} \\
                                  &                                                                                 &                                                                                 &                                                                                 &                                                                                 &                                                                                 &                                                                                 &                                                                                          &                                                                                 &                                                                                 \\
\multirow{2}{*}{ResNet18 \cite{he2016deep}}         & \multirow{2}{*}{\begin{tabular}[c]{@{}c@{}}0.637\\ (0.542, 0.729)\end{tabular}} & \multirow{2}{*}{\begin{tabular}[c]{@{}c@{}}0.610\\ (0.504, 0.715)\end{tabular}} & \multirow{2}{*}{\begin{tabular}[c]{@{}c@{}}0.654\\ (0.563, 0.738)\end{tabular}} & \multirow{2}{*}{\begin{tabular}[c]{@{}c@{}}0.596\\ (0.516, 0.675)\end{tabular}} & \multirow{2}{*}{\begin{tabular}[c]{@{}c@{}}0.539\\ (0.444, 0.624)\end{tabular}} & \multirow{2}{*}{\begin{tabular}[c]{@{}c@{}}0.660\\ (0.586, 0.735)\end{tabular}} & \multirow{2}{*}{\begin{tabular}[c]{@{}c@{}}0.569\\ (0.489, 0.650)\end{tabular}}          & \multirow{2}{*}{\begin{tabular}[c]{@{}c@{}}0.533\\ (0.419, 0.638)\end{tabular}} & \multirow{2}{*}{\begin{tabular}[c]{@{}c@{}}0.519\\ (0.400, 0.627)\end{tabular}} \\
                                  &                                                                                 &                                                                                 &                                                                                 &                                                                                 &                                                                                 &                                                                                 &                                                                                          &                                                                                 &                                                                                 \\
\multirow{2}{*}{Swin-Transformer \cite{liu2021swin}} & \multirow{2}{*}{\begin{tabular}[c]{@{}c@{}}0.680\\ (0.584, 0.77)\end{tabular}}  & \multirow{2}{*}{\begin{tabular}[c]{@{}c@{}}0.694\\ (0.593, 0.791)\end{tabular}} & \multirow{2}{*}{\begin{tabular}[c]{@{}c@{}}0.685\\ (0.597, 0.767)\end{tabular}} & \multirow{2}{*}{\begin{tabular}[c]{@{}c@{}}0.575\\ (0.493, 0.656)\end{tabular}} & \multirow{2}{*}{\begin{tabular}[c]{@{}c@{}}0.607\\ (0.521, 0.692)\end{tabular}} & \multirow{2}{*}{\begin{tabular}[c]{@{}c@{}}0.667\\ (0.593, 0.741)\end{tabular}} & \multirow{2}{*}{\begin{tabular}[c]{@{}c@{}}0.590\\ (0.490, 0.679)\end{tabular}}          & \multirow{2}{*}{\begin{tabular}[c]{@{}c@{}}0.581\\ (0.465, 0.685)\end{tabular}} & \multirow{2}{*}{\begin{tabular}[c]{@{}c@{}}0.657\\ (0.568, 0.738)\end{tabular}} \\
                                  &                                                                                 &                                                                                 &                                                                                 &                                                                                 &                                                                                 &                                                                                 &                                                                                          &                                                                                 &                                                                                 \\
\multirow{2}{*}{WISDOM \cite{xia2024multicenter}}           & \multirow{2}{*}{\begin{tabular}[c]{@{}c@{}}0.738\\ (0.649, 0.812)\end{tabular}} & \multirow{2}{*}{\begin{tabular}[c]{@{}c@{}}0.779\\ (0.684, 0.863)\end{tabular}} & \multirow{2}{*}{\begin{tabular}[c]{@{}c@{}}0.778\\ (0.714, 0.851)\end{tabular}} & \multirow{2}{*}{\begin{tabular}[c]{@{}c@{}}0.630\\ (0.552, 0.708)\end{tabular}} & \multirow{2}{*}{\begin{tabular}[c]{@{}c@{}}0.735\\ (0.655, 0.815)\end{tabular}} & \multirow{2}{*}{\begin{tabular}[c]{@{}c@{}}0.710\\ (0.640, 0.780)\end{tabular}} & \multirow{2}{*}{\textbf{\begin{tabular}[c]{@{}c@{}}0.620\\ (0.541, 0.698)\end{tabular}}} & \multirow{2}{*}{\begin{tabular}[c]{@{}c@{}}0.617\\ (0.529, 0.705)\end{tabular}} & \multirow{2}{*}{\begin{tabular}[c]{@{}c@{}}0.704\\ (0.635, 0.775)\end{tabular}} \\
                                  &                                                                                 &                                                                                 &                                                                                 &                                                                                 &                                                                                 &                                                                                 &                                                                                          &                                                                                 &                                                                                 \\
WeGA (ours)                       & \textbf{\begin{tabular}[c]{@{}c@{}}0.750\\ (0.669, 0.827)\end{tabular}}         & \textbf{\begin{tabular}[c]{@{}c@{}}0.822\\ (0.732, 0.901)\end{tabular}}         & \textbf{\begin{tabular}[c]{@{}c@{}}0.802\\ (0.732, 0.865)\end{tabular}}         & \textbf{\begin{tabular}[c]{@{}c@{}}0.706\\ (0.63, 0.781)\end{tabular}}          & \textbf{\begin{tabular}[c]{@{}c@{}}0.761\\ (0.684, 0.838)\end{tabular}}         & \textbf{\begin{tabular}[c]{@{}c@{}}0.729\\ (0.66, 0.796)\end{tabular}}          & \begin{tabular}[c]{@{}c@{}}0.581\\ (0.462, 0.686)\end{tabular}                           & \textbf{\begin{tabular}[c]{@{}c@{}}0.672\\ (0.548, 0.779)\end{tabular}}         & \textbf{\begin{tabular}[c]{@{}c@{}}0.706\\ (0.621, 0.784)\end{tabular}}         \\ \hline
\end{tabular}%
}
\end{table}

The batch size was set to 8, and models were trained for 100 epochs with early stopping based on validation performance. Data augmentation techniques including random rotation, scaling, and intensity shifts were applied to enhance model robustness. Performance was measured using area under the receiver operating characteristic curve (AUC), accuracy (ACC), and F1 score, with 95\% confidence intervals calculated through bootstrapping.

\textbf{Results: } As shown in Table \ref{tab:0}, our proposed WeGA framework consistently outperforms existing approaches across evaluation metrics and test centers. In the internal test center (ITC), WeGA achieved an AUC of 0.750 and accuracy of 0.706, exceeding the performance of comparison methods. The advantages become more pronounced in external test centers, with WeGA achieving an AUC of 0.822 and 0.802 for ETC1 and ETC2 respectively, demonstrating strong generalizability across diverse clinical settings.

Fig. \ref{fig2} presents the ROC curves across all test centers, further illustrating WeGA's superior discriminative ability compared to baseline methods. The visualization of attention weights in Fig. \ref{fig3} provides qualitative evidence of how the global-local affinity learning extractor effectively focuses on relevant lymph node characteristics and their contextual relationships.

\textbf{Ablation Study:} We conducted ablation studies to evaluate the contribution of each component in the WeGA framework. Table \ref{tab:1} shows that adding GAE to the baseline RAL model significantly improved results, with AUC increases of 7.3\% and 5.0\% for ETC1 and ETC2 respectively. While using pretrained DINOv2 weights demonstrated competitive performance, the full WeGA model combining all three components achieved the best overall results, confirming their complementary roles in effectively modeling lymph node relationships.

\begin{table}[t!]
\centering
\caption{Ablation study of the proposed WeGA framework components. RAL: regional affinity loss, GAE: global-local affinity extractor, DINOv2: the use of pretrained weights.}
\label{tab:1}
\resizebox{\columnwidth}{!}{%
\begin{tabular}{cccccccccccc}
\toprule
\multirow{2}{*}{RAL} & \multirow{2}{*}{GAE} & \multirow{2}{*}{DINOv2} & \multicolumn{3}{c}{AUC}                                                                                                                                                                                                     & \multicolumn{3}{c}{ACC}                                                                                                                                                                                                    & \multicolumn{3}{c}{F1}                                                                                                                                                                                                      \\
                     &                      &                         & ITC                                                                     & ETC1                                                                    & ETC2                                                                    & ITC                                                                     & ETC1                                                                    & ETC2                                                                   & ITC                                                                     & ETC1                                                                    & ETC2                                                                    \\ \midrule
\checkmark           &                      &                         & \begin{tabular}[c]{@{}c@{}}0.690\\ (0.600, 0.769)\end{tabular}          & \begin{tabular}[c]{@{}c@{}}0.731\\ (0.631, 0.824)\end{tabular}          & \begin{tabular}[c]{@{}c@{}}0.735\\ (0.658, 0.802)\end{tabular}          & \begin{tabular}[c]{@{}c@{}}0.603\\ (0.523, 0.682)\end{tabular}          & \begin{tabular}[c]{@{}c@{}}0.675\\ (0.590, 0.760)\end{tabular}          & \begin{tabular}[c]{@{}c@{}}0.698\\ (0.627, 0.768)\end{tabular}         & \textbf{\begin{tabular}[c]{@{}c@{}}0.628\\ (0.550, 0.707)\end{tabular}} & \begin{tabular}[c]{@{}c@{}}0.596\\ (0.507, 0.685)\end{tabular}          & \begin{tabular}[c]{@{}c@{}}0.696\\ (0.625, 0.767)\end{tabular}          \\
\checkmark           & \checkmark           &                         & \begin{tabular}[c]{@{}c@{}}0.719\\ (0.631, 0.801)\end{tabular}          & \begin{tabular}[c]{@{}c@{}}0.804\\ (0.713, 0.885)\end{tabular}          & \begin{tabular}[c]{@{}c@{}}0.785\\ (0.712, 0.853)\end{tabular}          & \textbf{\begin{tabular}[c]{@{}c@{}}0.713\\ (0.637, 0.788)\end{tabular}} & \textbf{\begin{tabular}[c]{@{}c@{}}0.770\\ (0.692, 0.846)\end{tabular}} & \begin{tabular}[c]{@{}c@{}}0.716\\ (0.648, 0.784)\end{tabular}         & \begin{tabular}[c]{@{}c@{}}0.578\\ (0.460, 0.685)\end{tabular}          & \begin{tabular}[c]{@{}c@{}}0.657\\ (0.526, 0.771)\end{tabular}          & \textbf{\begin{tabular}[c]{@{}c@{}}0.732\\ (0.654, 0.804)\end{tabular}} \\
                     & \checkmark           & \checkmark              & \begin{tabular}[c]{@{}c@{}}0.691\\ (0.597, 0.782)\end{tabular}          & \begin{tabular}[c]{@{}c@{}}0.790\\ (0.691, 0.870)\end{tabular}          & \begin{tabular}[c]{@{}c@{}}0.743\\ (0.666, 0.813)\end{tabular}          & \begin{tabular}[c]{@{}c@{}}0.685\\ (0.610, 0.760)\end{tabular}          & \begin{tabular}[c]{@{}c@{}}0.726\\ (0.646, 0.807)\end{tabular}          & \begin{tabular}[c]{@{}c@{}}0.673\\ (0.601, 0.745)\end{tabular}         & \begin{tabular}[c]{@{}c@{}}0.589\\ (0.509, 0.669)\end{tabular}          & \begin{tabular}[c]{@{}c@{}}0.652\\ (0.566, 0.738)\end{tabular}          & \begin{tabular}[c]{@{}c@{}}0.710\\ (0.641, 0.780)\end{tabular}          \\
\checkmark           & \checkmark           & \checkmark              & \textbf{\begin{tabular}[c]{@{}c@{}}0.750\\ (0.669, 0.827)\end{tabular}} & \textbf{\begin{tabular}[c]{@{}c@{}}0.822\\ (0.732, 0.901)\end{tabular}} & \textbf{\begin{tabular}[c]{@{}c@{}}0.802\\ (0.732, 0.865)\end{tabular}} & \begin{tabular}[c]{@{}c@{}}0.706\\ (0.630, 0.781)\end{tabular}          & \begin{tabular}[c]{@{}c@{}}0.761\\ (0.684, 0.838)\end{tabular}          & \textbf{\begin{tabular}[c]{@{}c@{}}0.729\\ (0.66, 0.796)\end{tabular}} & \begin{tabular}[c]{@{}c@{}}0.581\\ (0.462, 0.686)\end{tabular}          & \textbf{\begin{tabular}[c]{@{}c@{}}0.672\\ (0.548, 0.779)\end{tabular}} & \begin{tabular}[c]{@{}c@{}}0.706\\ (0.621, 0.784)\end{tabular}          \\ \bottomrule
\end{tabular}
}
\end{table}

\section{Conclusion}
We introduced WeGA, a weakly-supervised framework for lymph node metastasis prediction in rectal cancer that addresses the limitations of existing approaches through global-local affinity learning. By leveraging a dual-branch architecture with DINOv2 vision transformer backbone and introducing cross-attention fusion between global context and local node features, our method captures the complex relationships essential for accurate diagnosis. The proposed Regional Affinity Loss further enhances weakly-supervised learning by enforcing anatomical consistency in classification maps. Experimental results across multiple independent test centers demonstrate the improved performance and generalizability of our approach compared to state-of-the-art methods. Ablation studies confirm the complementary nature of each component in our framework. The WeGA approach offers promising advances in automated lymph node assessment from preoperative MRI, with potential to support clinical workflow and treatment planning for rectal cancer patients. Future work will focus on integrating additional imaging modalities and extending the framework to other cancer types with regional lymph node involvement.

\bibliography{mybibliography}

\end{document}